\pdfoutput=1

\documentclass[aip,jcp,reprint,floatfix,a4paper]{revtex4-1}  
\usepackage{cmap}
\usepackage[T1]{fontenc}
\usepackage{times}
\usepackage{helvet}
\usepackage{fixmath}
\DeclareSymbolFont{UPM}{U}{eur}{m}{n}
\DeclareMathSymbol{\partial}{0}{UPM}{"40}
\usepackage{graphicx}
\usepackage{amsmath}
\usepackage{geometry}
\usepackage[dvipsnames]{xcolor}
\usepackage{tikz,pgfplots}
\usetikzlibrary{backgrounds}
\tikzstyle{every picture}+=[remember picture]
\pgfplotsset{
tick label style={
/pgf/number format/.cd,
fixed,
fixed zerofill,
precision=1,
}
}
\pgfplotsset{compat=newest}
\pgfplotsset{
every tick label/.append style={
/pgf/number format/fixed, /pgf/number format/set thousands separator={}}
}

\usepgfplotslibrary{external}
{document}

\tikzsetexternalprefix{figures/}
\tikzset{external/system call={latex \tikzexternalcheckshellescape
-halt-on-error -interaction=batchmode -jobname "\image" "\texsource"; dvips -E
-o "\image".eps "\image".dvi; epstool --copy --bbox "\image".eps
"\image"new.eps; mv "\image"new.eps "\image".eps ;
 find ${PWD}/figures -maxdepth 1 -type f ! -iname
 "*.eps" -delete; }}

\usepackage[stretch=15,shrink=15,step=1]{microtype}
\SetProtrusion{encoding={*},family={*},series={*},size={6,7}}
              {1={ ,550},2={ ,300},3={ ,300},4={ ,300},5={ ,300},
               6={ ,300},7={ ,400},8={ ,300},9={ ,300},0={ ,300}}

\usepackage[nodayofweek]{datetime}
\newdateformat{myDate}{\THEDAY\ \monthname[\THEMONTH] \THEYEAR}

\newlength{\graphW}
\newlength{\graphH}
\usepackage{bm}
\usepackage{psfrag}
\usepackage{placeins}
\usepackage{csquotes}
\usepackage{mathtools}
\usepackage{color}
\usepackage{bbm}
\makeatletter \newcommand \listoftodos{\section*{Todo list} \@starttoc{tdo}}
\newcommand\l@todo[2]
 {\par\noindent \textbf{#2}:  #1\par\vspace{0.2cm}} \makeatother

\usepackage{hyperref}
\hypersetup{
        colorlinks=true,
        citecolor=black,
        filecolor=black,
        linkcolor=black,
        urlcolor=black
}

\begin{document}

\title[]{An enhanced version of the heat exchange algorithm
with excellent energy conservation properties.}

\author{P. Wirnsberger}
\email[Correspondence author. E-mail: ]{pw359@cam.ac.uk}

\author{D. Frenkel}%
\affiliation{ 
Department of Chemistry, University of Cambridge, Cambridge CB2 1EW, United
Kingdom%
}%

\author{C. Dellago}
\affiliation{%
Faculty of Physics, University of Vienna, 1090 Vienna, Austria \\
}%

\date{\today}%

\begin{abstract}
We propose a new algorithm for non-equilibrium molecular dynamics simulations of
thermal gradients. The algorithm is an extension of the heat exchange algorithm
developed by Hafskjold and co-workers [Mol. Phys. \textbf{80}, 1389 (1993); Mol. Phys. \textbf{81}, 251 (1994)], in which a certain amount of
heat is added to one region and removed from another by rescaling velocities appropriately. Since the amount of added and removed heat is the same and the dynamics between velocity rescaling steps is Hamiltonian, the heat exchange algorithm is expected to
conserve the energy.
However, it has been reported previously that the original version of the heat exchange algorithm exhibits a pronounced drift in the total
energy, the exact cause of which remained hitherto unclear.
Here, we show that the energy drift is due to the truncation error arising from
the operator splitting and suggest an additional coordinate integration
step as a remedy. The new algorithm retains
all the advantages of the original one whilst exhibiting excellent energy
conservation as illustrated for a Lennard-Jones liquid and SPC/E water.

\end{abstract}

\maketitle

\section{\label{sec:level1}Introduction}
Non-equilibrium molecular dynamics (NEMD) simulations allow us to study
transport phenomena and determine transport
coefficients~\cite{Evans1986}.
In studies of heat conduction, an external field is applied to the system
thereby driving it to a steady state.
The nature of the coupling between the external field and the system differs
between algorithms and determines whether a spatially homogeneous state
\cite{Evans1982}, a temperature gradient \cite{Ashurst1974, Baranyai1996, Ciccotti1980} or a heat flux is imposed
\cite{Ikeshoji1994, Muller-Plathe1997,Kuang2010, Kuang2012}. A suitable
algorithm for a particular application depends on its ability to
model the underlying physics correctly. If energy is supplied at a
constant rate in an experiment, for example, a thermostat which imposes a heat
flux would lend itself for the simulation. From a computational point of
view, generating a flux might be preferable, because it is
simpler to measure the temperature than the heat flux. 

One way to generate a heat flux in computer simulations involves swapping
kinetic energy between two subdomains of the simulation
box~\cite{Ikeshoji1994,Muller-Plathe1997}.
In the heat exchange (HEX) algorithm developed by Ikeshoji and
Hafskjold~\cite{Ikeshoji1994,Hafskjold1993}, a specific amount of heat is
periodically removed from one subdomain or reservoir, and supplied to the other. 
These two regions thus act as a heat sink and source, respectively.
The HEX method adjusts
the non-translational kinetic energy by velocity rescaling while preserving the
individual center of mass velocities of the two heat reservoirs.
Other methods use different procedures to generate heat fluxes.
In the reverse NEMD (RNEMD) method developed by
M\"uller-Plathe~\cite{Muller-Plathe1997}, the heat transfer is established by continuously identifying hot and cold particles inside the reservoirs and
exchanging their momenta.  Extensions of the RNEMD method were proposed by Kuang
and Gezelter~\cite{Kuang2010, Kuang2012}, who replaced the momenta swaps by velocity
rescaling moves. The velocity scaling and shearing (VSS) RNEMD
method~\cite{Kuang2012} allows for imposing a momentum flux in addition to the
thermal flux. However, we note that in the
absence of any momentum flux, the method is identical to the HEX algorithm.
Although these methods are widely applicable, they
all lack an attractive feature which is a formulation based on time continuous
equations of motion. Knowing the equations of motion is advantageous, for
example, if one is interested in studying system properties such as phase space
compressibility or the development of accurate integration schemes.

Due to its simplicity, the HEX algorithm is an attractive choice for simulating
a fluid in the absence of solid inhomogeneities. Since the same amount of
energy is added and removed, one would expect the algorithm to conserve the total energy exactly.
However, as pointed out in one of the original papers\cite{Hafskjold1993} and
subsequent work \cite{Bresme1996}, numerical implementations of the algorithm
lead to a considerable energy drift over simulation time scales of a few nanoseconds.
A change in total energy of several percent of the initial 
value was considered acceptable in past work.
Nevertheless, the energy drift is a severe restriction limiting the accessible
simulation time scales. Remedies to this 
problem either involve employing a smaller timestep or compensating the energy
drift with an additional thermostat~\cite{Bugel2008} which is undesirable,
because such thermostats may affect the very temperature profile that one aims to study.

In this work, we identify the underlying cause of the energy loss and
suggest a new algorithm to achieve improved energy
conservation. The paper is organised as follows: In Sec.~\ref{sec:hex}, we
summarise the HEX algorithm and its numerical implementation. We then show
that the integration scheme leads to an unphysical energy drift.
In order to understand the origin of this problem, we derive the 
equations of motion for continuous time in Sec.~\ref{sec:eom}. This allows us to
express the integration scheme as a Trotter factorisation of the Liouville
operator. In Sec.~\ref{sec:splitting}, we work out the leading-order error
term of the employed operator splitting. Based on our analysis, we propose an
enhanced algorithm in Sec.~\ref{sec:ehex} and compare the results in
Sec.~\ref{sec:results}.

\section{\label{sec:hex}HEX algorithm}
\begin{figure}[t]
  \centering
  \includegraphics{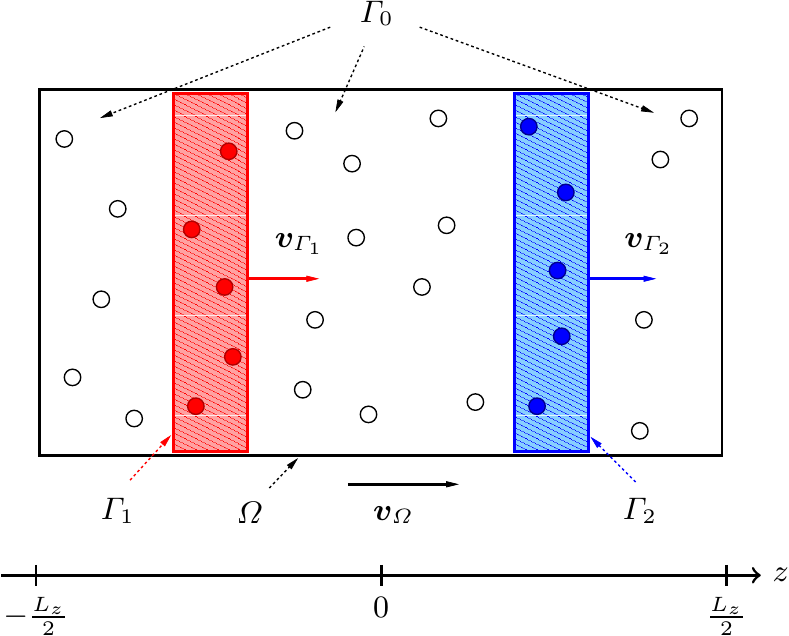}
  \caption{Illustration of the simulation box, $\Omega$, with
  Hamiltonian regions, $\Gamma_0$, a hot region, $\Gamma_1$
  (red), and a cold region, $\Gamma_2$ (blue).
           The centre of mass velocities of $\Omega$, $\Gamma_1$ and $\Gamma_2$ are $\mathbold v_\Omega$, $\mathbold v_{\Gamma_1}$ and $\mathbold v_{\Gamma_2}$, respectively.
           Atoms are represented by red/blue circles, if they are located in the
           hot/cold region and by empty circles otherwise.
            }
  \label{fig:boxhex}
\end{figure}

The goal of the HEX algorithm is to impose a constant heat
flux onto the system. This is accomplished by adding heat $\upDelta
Q_{\Gamma_k}$ at each timestep to $N_\Gamma$ pair-wise disjoint subdomains $\Gamma_k$, of the simulation box
 $\Omega$ (Fig.~\ref{fig:boxhex}).
Heat is subtracted if $\upDelta Q_{\Gamma_k}$ is negative.
We label those parts of the simulation box which 
are not thermostatted with $\Gamma_0$. 
The box contains $N$ atoms each labelled with a unique index.
If there is no net energy flux into the simulation box as we assume here, i.e.
$\sum_k \upDelta Q_{\Gamma_k} = 0$, the system will approach a steady state in
which heat fluxes are established between the subdomains.
The position and velocity vectors of atom $i$ are $\mathbold r_i$ and $\mathbold
v_i$, respectively.
Furthermore, we use $\mathbold v_{{\Gamma}_k}$ and $\mathbold v_{\Omega}$ to denote
the centre of mass velocities of the regions $\Gamma_k$ and the box $\Omega$,
respectively.

\subsection{\label{subsec:hex} Energy supply}

Energy is added or removed by rescaling the velocities
of all particles contained in region $\Gamma_k$ by the same factor $\xi_k$
and shifting them by a constant. The value of $\xi_k$ is chosen such that the non-translational kinetic energy of
that region,
\begin{equation}
{\mathcal K}_{\Gamma_k} = \sum_{{i \in \gamma_k}} \frac{ m_i v_i^2}{2} - \frac{m_{\Gamma_k} v_{\Gamma_k}^2}{2}
\label{eq:Ekinnt},
\end{equation}
changes by $\upDelta Q_{\Gamma_k}$ leaving $\mathbold v_{\Gamma_k}$ unchanged, where
$m_{\Gamma_k}$ is the total mass contained in $\Gamma_k$. 
The time-dependent index set $\gamma_k$ comprises all particles which are
located in $\Gamma_k$. Particles outside any thermostatted region are not
affected by this procedure.
For the individual region $\Gamma_k$, the velocity update can be formulated
as~\cite{Ikeshoji1994, Aubry2004}
\begin{equation}
\label{eq:hex}
\mathbold v_i \mapsto \bar{\mathbold v}_i = \xi_k \mathbold v_i +  \big(1-\xi_k\big) 
{\mathbold  v}_{\Gamma_k},
\end{equation}
where the rescaling factor is given by
\begin{equation}
\quad   \xi_k = \sqrt{ 1 + \frac{\upDelta Q_{\Gamma_k}}{ {\mathcal
K}_{\Gamma_k} } }.\label{eq:hexxii}
\end{equation}
Here, updated quantities are denoted with an overbar. 
It can easily be verified
that the update step given by Eq.~(\ref{eq:hex}) satisfies $\bar {\mathcal K}_{\Gamma_k}=   {\mathcal K}_{\Gamma_k}+ \upDelta Q_{\Gamma_k}$ 
and ${\bar {\mathbold v}}_{{\Gamma}_k} = \mathbold v_{{\Gamma}_k}$ . 
Since there is no net energy flux into the system according to our assumptions,
this also implies that the total system energy $E$ remains constant.

We note that the above formulation of the velocity update as presented by
Aubry~\textit{et al.}\cite{Aubry2004} is simpler than the one
which was originally proposed by Ikeshoji and Hafskjold~\cite{Ikeshoji1994}.
In the latter case, $\xi_k$ is a more complex function of the
velocities, but it is easy to see that both formulations are equivalent.

\subsection{\label{subsec:hexintegration}Time integration}
In order to keep track of the time evolution, it is convenient to introduce some
additional notation. We label all
quantities sampled at time $t=n \upDelta t$ with a superscript $n$, where $\upDelta t$ is the
timestep. In addition we define $\xi_0$ to be unity at all times and $k(\mathbold
r_i)$ to be the index of the region in which particle $i$ is located. The
current state of the system is fully described by a $6N$-dimensional vector $\mathbold
x=(\mathbold r, \mathbold v)$ in phase space, where the vectors $\mathbold r$ and
$\mathbold v$ contain all particle positions and velocities, respectively. 
The HEX algorithm for velocity Verlet~\cite{Swope1982a} can then be formulated as
\begin{subequations}
\label{eq:inthex}
\begin{alignat}{2}
               {\bar{\mathbold v}}_i^{n}  			&= && \quad \xi_{k({\mathbold
               r}_i)}^{n} {{\mathbold v}}_i^{n} +
               \Big(1-\xi_{k({\mathbold r}_i)}^{n}\Big) { {\mathbold
               v}}_{\Gamma_{k({\mathbold r}_i)}}^{n}, \\
               {\bar{\mathbold v}}_i^{n+\frac 12}  	&= && \quad {\bar{\mathbold
               v}}_i^n + \frac{\upDelta t}{2 m_i} {{\mathbold f}}_i^n,  \\
               {{\mathbold r}}_i^{n+1}   				&= && \quad {{ \mathbold r}}_i^n +
               \upDelta t \ {\bar{\mathbold v}}_i^{n+\frac 12},  \\
               \mathbold f_i^{n+1}           			&= && \quad -\nabla_{\mathbold r_i}
               \ U(\mathbold r)  \left. \right|_{\mathbold r = {\mathbold r}^{n+1}},
               \\
               {\bar{\mathbold v}}_i^{n+1}      		&= && \quad {\bar{\mathbold
               v}}_i^{n+\frac 12} + \frac{\upDelta t}{2 m_i} \mathbold f_i^{n+1}, 
               \\ 
                   {{\mathbold v}}_i^{n+1}  			&= && \quad \bar{{
                   \xi}}_{k({\mathbold r}_i)}^{n+1} {\bar{\mathbold v}}_i^{n+1}
                   + \Big(1-\bar{{\xi}}_{k({\mathbold r}_i)}^{n+1}\Big)
                   {\bar{\mathbold v}}_{\Gamma_{k({\mathbold r}_i)}}^{n+1},
\end{alignat}
\end{subequations}
where $U(\mathbold r)$ is the potential energy and $\mathbold f_i$ the force acting
on particle $i$.
For the entire scheme to be symmetric, half the energy is supplied at the
beginning of the timestep and the other half at the end. The scaling factors
$\xi_{k({\mathbold  r}_i)}^{n}$ and $\bar{{\xi}}_{k({\mathbold r}_i)}^{n+1}$ are
evaluated using Eq.~(\ref{eq:hexxii}) at the states $\big(\mathbold r^{n}, {
{\mathbold v}}^{n}\big)$ and $\big(\mathbold r^{n+1}, {\bar
{\mathbold v}}^{n+1}\big)$, respectively.

We note that in the original work~\cite{Hafskjold1993,Ikeshoji1994}, the
authors do not provide any details about when exactly the thermostatting step
should happen. For comparison, we also tested an asymmetric version of the
algorithm (HEX/a), where all the energy is supplied at the end of the timestep.
In this case, the initial velocity update reduces to the identity operation.

\subsection{\label{subsec:hexmodel}Model system}
We studied the energy conservation of the HEX algorithm for a Lennard-Jones
(LJ) fluid using the simulation package LAMMPS (version 9Dec14)~\cite{Plimpton1995}. 
The symmetric, pairwise LJ potential is given by~\cite{Frenkel2002}
\begin{align}
u_\text{LJ}(r) = 4 \epsilon \left[ {\left(\frac{\sigma}{r}\right)}^{12} - {\left(\frac{\sigma}{r}\right)}^{6} \right], \label{eq:lj}
\end{align}
where $\epsilon$ is the depth of the potential and $\sigma$ the effective atomic diameter. 
In order to rule out any effects due to simple spherical truncation of the
potential, we employed a slightly modified potential which is given by~\cite{Toxvaerd2011}
\begin{align}
\label{eq:ljshifted}
{u}_\text{SF} (r) = u_\text{LJ}(r) -  u_\text{LJ}(r_\text{s})  - (r - r_\text{s}) u^\prime_\text{LJ}(r_\text{s})
\end{align}
for $r \leq r_\text{s}$ and zero otherwise, where $r_\text{s}$ is the
cutoff.
From the functional form of Eq.~(\ref{eq:ljshifted}) it is clear that $u_\text{SF}(r)$ and $u^\prime_\text{SF}(r)$ are both continuous
at the cutoff. We employed a value of $r^*_\text{s} = 3$ for all
simulations in this section. (Reduced quantities are
labelled with an asterisk.)

\subsection{\label{subsec:hexequi}Equilibration}
The rectangular simulation box with dimensions $L^*_z/2 = L^*_x  = L^*_y =
10.58$ comprised $N=2000$ atoms resulting in a density of $\rho^* = 0.8444$. 
The thermodynamic conditions considered
in this work are similar to those in Ref.~\onlinecite{Ikeshoji1994}.
Starting from an initial lattice structure with zero linear momentum, the system was heated up to twice the target temperature of $T^*=0.72$
and subsequently cooled down again at the same rate. The thermostatting during
this initial period was achieved by velocity rescaling and the entire annealing process took $2.5\times10^4$ timesteps. 
The equations of motion were integrated with the velocity Verlet algorithm using
a timestep of ${\upDelta t^*=0.002}$.
We then increased the timestep to $\upDelta t^* =0.004$ and
carried out a $2\times10^5$ timestep \textit{NVT} simulation using a
Nos\'e--Hoover thermostat~\cite{Nose1984, Hoover1985} with a
relaxation time of $\tau^* = 0.5$. During this run we computed the average system energy.
Using the HEX algorithm, we then adjusted the energy of the last configuration
and used it as input for another $2\times10^5$ timestep \textit{NVE}
equilibration run. This procedure allowed us to achieve an average equilibrium
temperature of ${T^* = (0.7200 \pm 0.0002)}$. The error bar corresponds
to one standard deviation of the error of the mean, the variance of which was
estimated using block average analysis~\cite{Frenkel2002}.

As a reference for the energy conservation in equilibrium, we 
carried out an additional set of
\textit{NVE} simulations at various timesteps. 
With the above protocol we matched the
temperature of these runs to be close to the one inside the hot reservoir
in the NEMD case. The average temperature was ${T^*= (0.8400\pm0.0002)}$.

\subsection{\label{subsec:hexenergy}Energy conservation during NEMD}
The previously equilibrated structures were subjected to a temperature
gradient along the $z$-axis using the HEX algorithm.  Always starting from the
same phase space point, we varied the timestep for a fixed energy flux
$\mathcal F_{\Gamma_k} = \upDelta Q_{\Gamma_k}/\upDelta t$ into the reservoir. The two
thermostatted regions 
are centred at the points $z = \pm L_z/4$ and have a width of 2 in reduced
units~(Fig.~\ref{fig:boxhex}).
During each timestep, the heat $\upDelta Q$ is taken from $\Gamma_2$ and added to
$\Gamma_1$ ($\upDelta Q_{\Gamma_1} = -\upDelta Q_{\Gamma_2} = \upDelta Q > 0$).
We waited for 100 reduced time units for any transient behaviour to
disappear and to allow the system to reach a steady state. The production
run of 5000 reduced time units started at $t^*=0$.

In order to capture the spatial variation of the temperature, 
we divided the $z$-axis into $N_\text{b}$ bins.
We use the notation $X_j$ for the evaluation of
a quantity $X$ over bin $j$ and assign the value to
the centre of the bin. The instantaneous kinetic temperature of bin $j$
is then given by
\begin{equation}
T_j = \frac{2\mathcal K_j}{ (N_j f -3) k_\text{B} },
\end{equation}
where $\mathcal K_j$ is the total non-translational kinetic energy of the bin,
$N_j$ the number of atoms contained in the bin and $k_\text{B}$ Boltzmann's constant. The
quantity $f$ is the number of degrees of freedom per atom
($f_\text{LJ}=3$ and $f_\text{SPC/E} = 2$). 
We subtracted three degrees of freedom to account for the centre of mass
velocity of the bin.
In the stationary state, the heat flux between the reservoirs in
Fig.~\ref{fig:boxhex} is given by
\begin{equation}
J_{Q,z} = \frac{\upDelta Q}{2\upDelta t L_x L_y} = \frac{\mathcal F}{2 L_x L_y},
\end{equation}
where the factor of 2 in the denominator accounts for the periodic setup. 
Considering a reference layer, this is intuitively clear, 
because half the supplied heat will flow to the other reservoir
in the reference box and the other half to its image in the neighbouring box. 
The heat flux is an input parameter of the HEX algorithm, which we set to $0.15$
in reduced units.
\begin{figure}[t]
   \centering
   \includegraphics{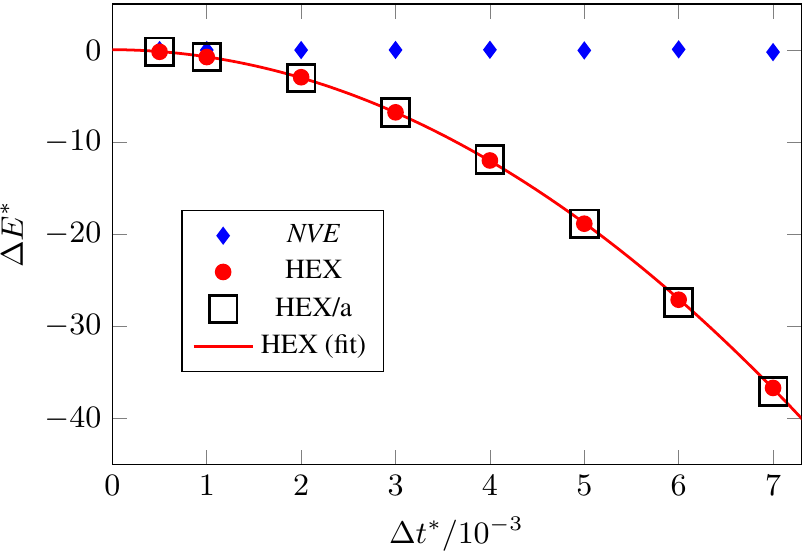}
   \caption{
	Energy loss for LJ at the final time $t^*=5000$ as a function of the timestep.
	Each point in the figure corresponds to a separate simulation. The
	equilibrium run (blue diamonds) is compared to the symmetric (red circles) and asymmetric (black
	squares) versions of the HEX algorithm, respectively. A quadratic fit (red,
	solid line) was carried out for the symmetric version.
   }
   \label{fig:dE_lj}
\end{figure}

The dependence of the energy loss at the final time on the timestep is shown in
Fig.~\ref{fig:dE_lj}.
From the quadratic fit, it is clear that the HEX algorithm exhibits an
energy drift which scales as $\mathcal O({\upDelta t}^2)$. On the other hand, 
the energy was conserved perfectly well in \textit{NVE} simulations
 at the peak temperature inside the hot reservoir. (The temperature profiles are
discussed in Sec.~\ref{sec:results}.)

\section{\label{sec:eom} Equations of motion}
To gain a better understanding of the energy drift of the HEX algorithm, we
first derive the ordinary differential equations (ODEs) solved by the algorithm
in the limit $\upDelta t \to 0$.
To this end we consider the velocity update for continuous time. Dropping all particle and region indices for readability and eliminating the intermediate velocities, we 
can cast Eq.~(\ref{eq:inthex}f) into
\begin{alignat}{2}
  {{\mathbold v}}^{n+1} &=&& \quad {\bar \xi^{n+1}} \left[{\xi^{n}} {{{\mathbold
  v}}}^{n} + \left(1-{\xi^{n}}\right) {\mathbold
  v}_{\Gamma}^{n} + \frac{\upDelta t}{2 m} \left(\mathbold f^{n} + \mathbold
  f^{n+1} \right) \right] \nonumber \\
   & && \quad +\left(1-{\bar \xi^{n+1}}\right) {\bar{\mathbold
   v}}_{\Gamma}^{n+1}.\label{eq:hexvcont1} 
\end{alignat}
If we subtract ${\mathbold v}^n$ on both sides and divide by the timestep, 
we get
\begin{alignat}{2}
  \frac{{{\mathbold v}}^{n+1}- {\mathbold v}^{n} }{\upDelta t} 
  & = && \quad
  \frac{{\bar\xi}^{n+1}}{2 m} \left(  \mathbold f^{n} +
  \mathbold f^{n+1} \right) \\
  & + && \quad
  \frac{\left( {\bar \xi }^{n+1} {\xi }^{n}-1\right) {
  {\mathbold v}}^{n} }{\upDelta t}\nonumber \\
  & + && \quad 
  \frac{{\bar \xi }^{n+1}\left(1- {\xi }^{n}\right)
  {{\mathbold v}}_{\Gamma}^{n} +\left(1- {{\bar \xi} }^{n+1}\right)
  {\bar {\mathbold v}}_{\Gamma}^{n+1}     } {\upDelta t} . 
 \nonumber
\end{alignat}
It is straightforward to show that 
\begin{equation*}
   \frac{  \left( {\bar \xi }^{n+1} {\xi }^{n} -1  \right)           
           { {\mathbold v}}^{n} }    {\upDelta t} 
        \to 
        \frac{ \mathcal F_{\Gamma} {\mathbold v}(t^n)}{2 \mathcal K_\Gamma(t^n)}
\end{equation*}
and
\begin{equation*}
     \frac{{\bar \xi }^{n+1}\left(1- {\xi }^{n}\right)
  {{\mathbold v}}_{\Gamma}^{n} +\left(1- {{\bar \xi} }^{n+1}\right)
  {\bar {\mathbold v}}_{\Gamma}^{n+1}     } {\upDelta t} 
  \to -\frac{ \mathcal F_{\Gamma} {\mathbold v}_{\Gamma}(t^n)}{2
  \mathcal K_\Gamma(t^n)}  
\end{equation*}
in the limit of $\upDelta t \to 0$. From Eq.~(\ref{eq:inthex}c), it 
is immediately obvious that the derivative of the coordinates is given by the
velocities. \\

The continuous equations of motion solved by
the HEX algorithm are therefore given by
\begin{subequations}
\begin{align}
  \dot{\mathbold r}_i &= \mathbold v_i, \label{eq:hexvcont3a} \\
  \dot{\mathbold v}_i &= \frac{\mathbold f_i}{m_i} + \frac{\mathbold \eta_i}{m_i},
  \label{eq:hexvcont3b}
\end{align}
\end{subequations}
where the thermostatting force is defined as
\begin{equation}
  \label{eq:eta}
  \mathbold \eta_i = 
  \begin{cases} m_i   \frac{\mathcal F_{\Gamma_{k(\mathbold r_i)}}}{2 \mathcal
  K_{\Gamma_{k(\mathbold r_i)}}} \left(\mathbold v_i - \mathbold v_{\Gamma_{k(\mathbold
  r_i)}} \right) &\mbox{if $k(\mathbold r_i) > 0$,} \\
   0 &\mbox{otherwise. } 
  \end{cases}
\end{equation}
In order for the equations to be well-defined, we
assume that there are sufficiently many particles inside any
reservoir, i.e. regions with $k(\mathbold r_i) >	0$, such that the
non-translational kinetic energy never vanishes. Outside the reservoirs the thermostatting force
is zero and the particles obey Hamiltonian motion. Some further properties of
the equations are analysed in Appendix~\ref{secapp:hex}.

\section{\label{sec:splitting} Operator splitting}
Our goal is to show that the energy drift is caused by
higher-order truncation terms, which are not taken into
account in the time integration. These terms can be derived easily once
the integration scheme is expressed in terms of a Trotter factorisation of the
Liouville operator.

\subsection{\label{subsec:trotter} Trotter factorisation}
Tuckerman \textit{et al.}~\cite{Tuckerman1992} showed that reversible
integrators can be generated based on a Trotter factorisation of the Liouville
operator $iL$.
Utilising the same theoretical framework, we consider the splitting
\begin{subequations}
\label{eq:hexvcont2}
\begin{align}
i L   &= iL_1 + iL_2, \\
iL_1 &= \sum_{j=1}^N \sum_{\alpha \in \{x,y,z\}} \frac{\eta_{j,\alpha}}{m_j}  
  \frac{\partial}{\partial v_{j,\alpha}}, \\
iL_2 &= \sum_{j=1}^N  \sum_{\alpha \in \{x,y,z\}} \left[ \frac{f_{j,\alpha}}{m_j} 
\frac{\partial}{\partial v_{j,\alpha}} + v_{j,\alpha} \frac{\partial}{\partial
r_{j,\alpha}} \right]
\end{align}
\end{subequations}
and apply it to the current state of the system which is fully described by
$\mathbold x$ in the $6N$-dimensional phase space. The exact time evolution of the
system is formally given by
 \begin{equation}
\label{eq:xexact}
\mathbold x_\text{ex}(t) = \text{e}^{t iL } \mathbold x(0).
\end{equation} 
Unfortunately, it is not feasible to
evaluate this expression analytically for the equations we are interested in.
The problem can be simplified, however, by considering the approximation
\begin{equation}
\label{eq:hexsplit}
\mathbold x(t) = {\left[\text{e}^{\frac{\upDelta t}{2} iL_1 }
                      \text{e}^{\upDelta t iL_2 }
                      \text{e}^{\frac{\upDelta t}{2} iL_1 }\right]}^{P} \mathbold
                      x(0),
\end{equation}
where $P$ is an integral number which implicitly defines the timestep through
$\upDelta t = t/P$.
The operator $\text{e}^{\frac{\upDelta
t}{2} iL_1}$ acts on the velocities and adds
the energy $\upDelta Q/2$ to the system. In fact, as shown in
Appendix~\ref{secapp:hex}, the velocity update of the HEX
algorithm is the exact solution of this operation.
Hamilton's equations of motion are then integrated with $\text{e}^{\upDelta t
iL_2}$ followed by the second energy supply. For the analysis in the next
subsection, we assume that all operations in Eq.~(\ref{eq:hexsplit}) can be carried
out analytically, although in the simulation we use an additional approximation
of $\text{e}^{\upDelta t iL_2}$, as discussed in Sec.~\ref{sec:ehex}.

\subsection{\label{subsec:trunc} Local truncation error}
The splitting given by Eq.~(\ref{eq:hexsplit}) is known as Strang
splitting~\cite{Strang1968}. It has the local truncation
error~\cite{LeVeque1983}
\begin{equation}
\label{eq:strang}
\mathbold x(\upDelta t) - \mathbold x_\text{ex}(\upDelta t) = {\upDelta t}^3 \mathcal E
 \mathbold x_\text{ex}(0) + \mathcal O\left({\upDelta t}^4\right),
\end{equation}
where the first term on the RHS is determined by the
operator
\begin{equation}
\label{eq:strang2}
\mathcal E  = \frac {1}{12} \big[ iL_2, \left[ iL_2, iL_1 \right] \big] 
              -\frac {1}{24} \big[ iL_1, \left[ iL_1, iL_2 \right] \big]
\end{equation}
and $\left[A,B\right] = AB - BA$ is the commutator. Rearranging terms, we find
\begin{equation}
\label{eq:strang3}
\mathbold x(\upDelta t) -{\upDelta t}^3 \mathcal E
 \mathbold x_\text{ex}(0) = \mathbold x_\text{ex}(\upDelta t) +   \mathcal
O\left({\upDelta t}^4\right).
\end{equation}
This means that the key to improving the accuracy of the numerical approximation is to apply the correction 
$- {\upDelta t}^3 \mathcal E\mathbold x_\text{ex}(0)$ to the original solution. 
Alternatively, we can also use a correction $- {\upDelta t}^3 \mathcal E
\tilde {\mathbold x}(\upDelta t)$, where $\tilde{\mathbold x}(\upDelta t) =
\mathbold x_\text{ex}(0) + \mathcal O(\upDelta t)$, without changing the order
of the truncation error.

\section{\label{sec:ehex} Enhanced HEX algorithm}
The analysis of the previous section remains valid for any approximation of
$\text{e}^{\upDelta t iL_2}$ which is sufficiently accurate.
This is necessarily the case if the local truncation error is $\mathcal
O\left({\upDelta t}^4\right)$ or higher. Velocity Verlet
integration is less accurate than that and has a local truncation error of
$\mathcal O\left({\upDelta t}^3\right)$. 
Nevertheless, we found that it is fully
sufficient to consider a coordinate correction of the form of
Eq.~(\ref{eq:strang3}) to get hold of the energy loss. We therefore ignored the
additional velocity Verlet truncation error and all other correction terms in
Eq.~(\ref{eq:strang3}) affecting velocities only.

This analysis leads us directly to the 
\textit{enhanced heat exchange (eHEX)} algorithm, which is defined through the
update sequence
\begin{subequations}
\label{eq:intehex}
\begin{alignat}{2}
               {\bar{\mathbold v}}_i^{n}  &= && \quad \xi_{k({\mathbold
               r}_i)}^{n} {{\mathbold v}}_i^{n} +
               \Big(1-\xi_{k({\mathbold r}_i)}^{n}\Big) { {\mathbold
               v}}_{\Gamma_{k({\mathbold r}_i)}}^{n}, \\
               {\bar{\mathbold v}}_i^{n+\frac 12}  &= && \quad {\bar{\mathbold
               v}}_i^n + \frac{\upDelta t}{2 m_i} {{\mathbold f}}_i^n,  \\
               {\bar{\mathbold r}}_i^{n+1}   &= && \quad {{ \mathbold r}}_i^n +
               \upDelta t \ {\bar{\mathbold v}}_i^{n+\frac 12},  \\
               \mathbold f_i^{n+1}           &= && \quad -\nabla_{\mathbold r_i} \
               U(\mathbold r)  \left. \right|_{\mathbold r = {\bar {\mathbold
               r}}^{n+1}},
               \\
               {\bar{\mathbold v}}_i^{n+1}      &= && \quad {\bar{\mathbold
               v}}_i^{n+\frac 12} + \frac{\upDelta t}{2 m_i} \mathbold f_i^{n+1}, 
               \\ 
                   {{\mathbold v}}_i^{n+1}  &= && \quad { {\bar \xi}}_{k(\bar
                   {\mathbold r}_i)}^{n+1} {\bar{\mathbold v}}_i^{n+1} +
                   \Big(1-{\bar \xi}_{k(\bar{\mathbold r}_i)}^{n+1}\Big)
                   {\bar{\mathbold v}}_{\Gamma_{k(\bar{\mathbold r}_i)}}^{n+1}, \\
         {{\mathbold r}}_i^{n+1}   &= && \quad {\bar {\mathbold r}}_i^{n+1} -
         {\upDelta t}^3 \mathcal E {\bar {\mathbold r}}_i^{n+1}.                         
\end{alignat}
\end{subequations}
Apart from the last integration step and some relabelling, this scheme is
identical to the HEX algorithm. As shown in Appendix~\ref{secapp:hex}, the correction term is given by 
\begin{alignat}{2}
\label{eq:ehexrj}
 \mathcal E  {r}_{i,\alpha}  &= &&\frac{\eta_{i,\alpha}}{m_i
 \mathcal K_{\Gamma_{k(\mathbold r_i)}}} \Bigg[\frac{\mathcal
 F_{\Gamma_{k(\mathbold r_i)}}}{48} + \frac 16 \sum_{j\in \gamma_{k({
 {\mathbold r}}_i)}} \mathbold f_j \cdot \Big(\mathbold v_j - \mathbold
 v_{\Gamma_{k({ {\mathbold r}}_i)}}\Big) \Bigg] \nonumber \\
 & &&- \frac{\mathcal F_{\Gamma_{k(\mathbold r_i)}}}{12 \mathcal
 K_{\Gamma_{k(\mathbold r_i)}}} \Bigg[
 \frac{f_{i,\alpha}   }{m_i} - \frac{1}{m_{\Gamma_{k({{\mathbold r}}_i)}}}
 \sum_{j\in\gamma_{k({{\mathbold r}}_i)}} f_{j,\alpha}  \Bigg]
\end{alignat}
and evaluated at the state ${\bar {\mathbold x}}^{n+1}$.
We note that this expression vanishes for particles outside any reservoir,
because the thermostatting force is zero in that case. The
scaling factors $\xi_{k({\mathbold r}_i)}^{n}$ and ${\bar{\xi}}_{k(\bar{\mathbold r}_i)}^{n+1}$ are calculated at the system states $\mathbold x^n$ and $\bar{\mathbold x}^{n+1}$, respectively. 
As in the formulation of the original algorithm, we also consider the case where
all the energy is supplied asymmetrically in Eq.~(\ref{eq:intehex}f). We refer to this version of
the algorithm as eHEX/a.

\subsection{\label{subsec:rigid} Rigid molecules}

Employing constraining forces, we can extend
the eHEX algorithm to a system of rigid bodies, such as SPC/E
water~\cite{Berendsen1987}.
In the SHAKE algorithm~\cite{Ryckaert1977}, originally 
devised for Verlet integration, rigidity is imposed by solving
iteratively for a set of Lagrange multipliers.
If the underlying equations are integrated with the velocity Verlet algorithm, 
a second set of constraining forces is required
to eliminate velocity components along any fixed bond. This is taken
into account by the RATTLE algorithm \cite{Andersen1983} which we implemented in
LAMMPS.

To be compatible with the treatment of constraining forces in
LAMMPS, we consider the eHEX/a algorithm for rigid bodies. We use RATTLE to ensure that the
velocities and positions are satisfied up to the target tolerance after the
second velocity update (Eq.~(\ref{eq:intehex}e)). Provided that all sites of a
reference molecule are located in the same region, the scaling and shifting in
Eq.~(\ref{eq:intehex}f) does not violate the constraints.
For this reason, we only rescale an individual site of a molecule if its centre
of mass is located within the reservoir.
 
For the small fraction of molecules inside a reservoir, the coordinate
correction in Eq.~(\ref{eq:intehex}g) introduces an $\mathcal O({\upDelta t}^3)$
error in the bond distances. This error is small and of the same order as the
local error of the RATTLE algorithm itself~\cite{Andersen1983}.
For this reason, we consider an unconstrained update acceptable. However, we
monitored the maximum relative errors throughout all simulations. The constraining forces for the coordinates are 
recalculated at the end of the timestep to ensure that
the positions are correct after the subsequent velocity Verlet update. 
 
\subsection{\label{subsec:spce} Model system}
In addition to the monatomic system, we tested the eHEX/a algorithm for the
 SPC/E water model.
The simulation box with dimensions ${L_z/2=L_x=L_y=25.26~\text{\AA}}$
contained $1024$ molecules resulting in a density of
$0.95~\text{g/}\text{cm}^3$.
We used a real-space cutoff of $11~\text{\AA}$ for the
LJ and Coulomb interactions, which were evaluated with standard Ewald
summation~\cite{Frenkel2002}. The damping parameter was $\alpha  = 6.816/L_x$ 
with $9841$ $\mathbold k$-vectors (before employing symmetry properties of the
reciprocal sum).

Starting from a lattice structure, we rescaled velocities for $10$~ps to drive
the system close to a target temperature of 400~K.
We employed a timestep of 1~fs and the equations were integrated with velocity
Verlet. This was followed by a $2\times10^5$ timestep \textit{NVT} simulation
using a Nos\'e--Hoover thermostat with a relaxation
time of 1~ps.	
The total energy of the last configuration was then adjusted such that it 
corresponds to the average of the \textit{NVT} run.
The average temperature over a subsequent $2\times10^5$ timestep
\textit{NVE} simulation was ${(400.5\pm 0.2)~\text{K}}$. 

We then switched on the thermostat and waited for 100~ps for the system to
reach a steady state before starting with the 1~ns production run. 
The reservoirs of width $4~\text{\AA}$ were centred at the points $z=\pm
L_z/4$ and we imposed a heat flux of $4.08\times10^{10}~\text{W}/{\text{m}^2}$.
As a reference for the energy conservation, we carried out an additional set of 1~ns \textit{NVE} simulations at various
timesteps. The temperature in these runs was ${(468.0 \pm 0.2)~\text{K}}$.

\section{\label{sec:results} Results}
\begin{figure}[t]
   \centering
    \includegraphics{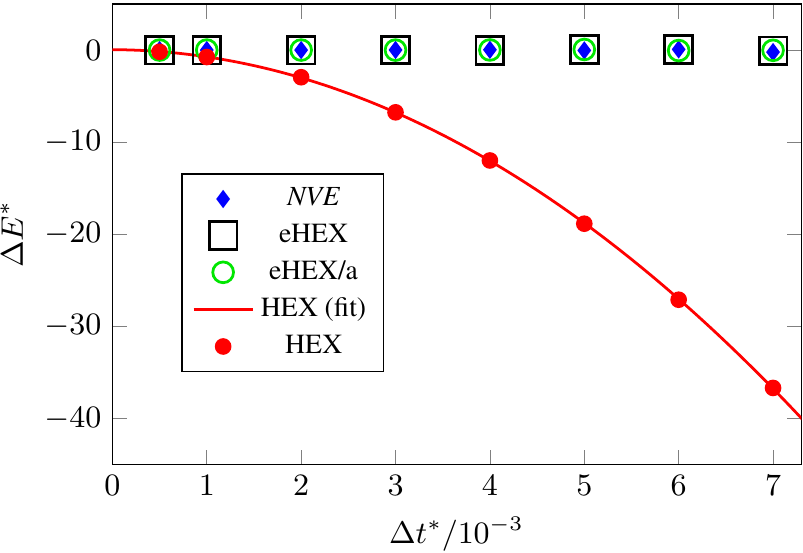}
   \caption{Energy loss for LJ at the final time $t^*=5000$ for various timesteps.
	The equilibrium run (blue diamonds) and the symmetric (black squares) and
	asymmetric (green, open circles) versions of the eHEX algorithm, respectively,
	do not show any appreciable drift. The energy loss of the HEX algorithm (red, full circles) together with a quadratic fit (red, solid line)
	is shown for comparison.
	}
   \label{fig:dEdt_lj2}
\end{figure}

\begin{figure}[t]
   \centering
   \includegraphics{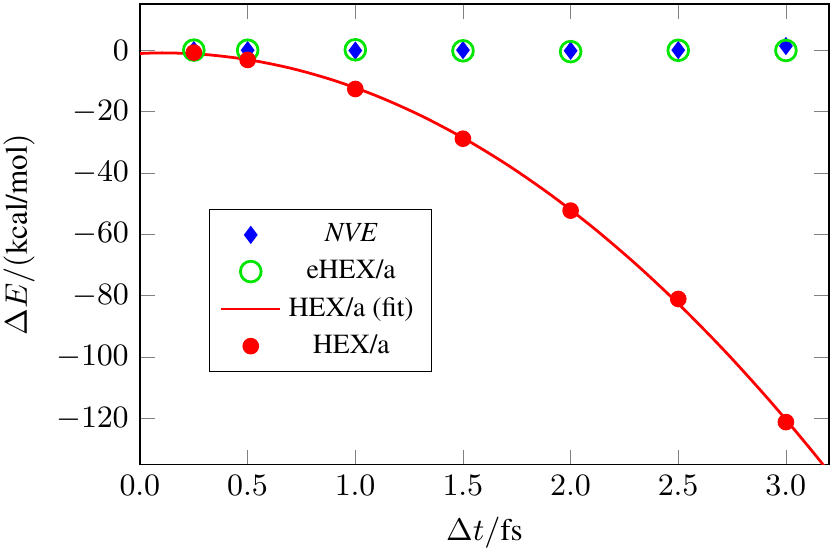}
\caption{Energy loss for SPC/E water at the final time $t=1~\text{ns}$ for
various timesteps. The equilibrium run (blue diamonds) is compared
to the asymmetric eHEX algorithm (green, open circles) and the
asymmetric HEX algorithm (red, full circles) together with a quadratic fit (red, solid line).}
\label{fig:dEdt_spce}
\end{figure}

The effect of the additional coordinate integration in the eHEX algorithm on the
total energy conservation is shown in
Figs.~\ref{fig:dEdt_lj2}--\ref{fig:dEdt_spce}.
As can be seen, the new algorithm exhibits excellent energy conservation. Even
for large timesteps close to the stability limit for a \textit{NVE}
simulation at the peak temperature (${\upDelta t}^*_\text{max, LJ} \approx 0.0075$
and ${\upDelta t}_\text{max, SPC/E} \approx 3.5~\text{fs}$),
there is no noticeable drift on this scale.
The energy loss of the HEX algorithm, on the other hand, is substantial. At the
largest timestep, the total system energy changed by about
$0.45\text{\%}$ for LJ and $1.6\text{\%}$ for SPC/E,
respectively.
Although an energy loss of several percent was considered acceptable in the past~\cite{Bresme1996,
Ikeshoji1994, Hafskjold1993}, it sets an upper limit to the accessible
simulation time scales.
The only way to circumvent this problem apart from coupling the system to an
additional thermostat is to decrease the timestep and thereby waste valuable computing time.
Based on a series of eight simulations at the largest timestep and with
different initial conditions, we can give a conservative estimate
of the improvement due to the new algorithms. For LJ we found that the eHEX
algorithm loses at least $500$ times less energy than the HEX algorithm ($450$ for eHEX/a as compared to HEX/a). For SPC/E water we found that the
eHEX/a algorithm improves the energy conservation by at least a factor of $100$ as compared to the HEX/a
algorithm. The accessible simulation time scale therefore increased by
two orders of magnitude.

\begin{figure}[t]
   \centering
   \includegraphics{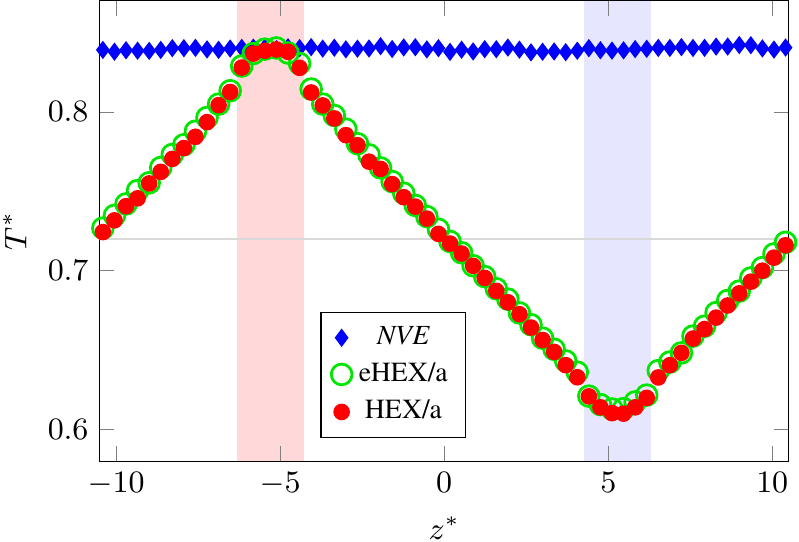}
\caption{Comparison of the temperature profiles for LJ.}
\label{fig:T_lj}
\end{figure}

\begin{figure}[t]
   \centering
   \includegraphics{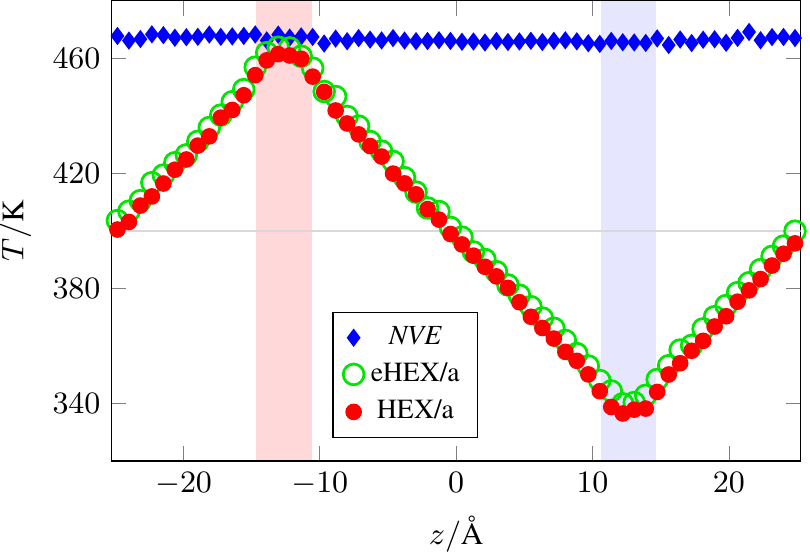}
\caption{Comparison of the temperature profiles for SPC/E water.}
\label{fig:T_spce}
\end{figure}

The spatial variation in temperature is shown in
Figs.~\ref{fig:T_lj}--\ref{fig:T_spce}. For the monatomic system, the results
agree well without any marked differences (Fig.~\ref{fig:T_lj}). We
note that in both cases there are some visible discontinuities in the
vicinity of the reservoirs. This should not be very surprising, since the
thermostatting force is also discontinuous. We found
that the gap decreases as we go to lower temperature gradients, because the
fluid can dissipate the heat sufficiently fast. A possible way of controlling
this gap is to employ a weight function and to redefine temperature such that it is continuous at the boundary of the reservoir~\cite{Baranyai1996}. 
This procedure allows for better control and is numerically convenient, but it
is not obvious which weight function is physically most meaningful.
Furthermore, generalising this approach to arbitrary reservoir shapes is challenging, because it requires some sort of
signed distance information to the boundary.

For SPC/E water the energy loss at large timesteps is reflected in a
slight drop of temperature (Fig.~\ref{fig:T_spce}). The overall profiles agree
well, but they are shifted by a few Kelvin.
This shift is consistent with the energy loss of about $1\%$ for the
$2.5~\text{fs}$ timestep. There
are no visible temperature discontinuities in the vicinity of the reservoirs. This
might be related to the fact that in our scheme the boundaries are naturally
smeared out as we only rescale entire molecules which could be intersected by
the reservoir boundary.

Although we omitted a constrained coordinate
update in the eHEX/a algorithm, the relative deviation
from the ideal bond distance never exceeded $1.1\times10^{-5}$. This was the
case for the largest timestep of 3~fs, but the error decays rapidly (with
${\upDelta t}^3$) such that it reduced to $3.6\times10^{-7}$ for a timestep of
1~fs.
The maximum induced relative velocity along any rigid bond was an order of magnitude lower
for both timesteps, respectively. Only a small fraction of
molecules inside a reservoir ($\approx 16\%$) suffers from
this inconsistency. We consider this error acceptable and an unconstrained
update justified. An extension of the eHEX algorithm to a constrained update is
possible in case higher precision is required. 

With regard to conservation of total linear momentum, we found that both
algorithms satisfied this condition perfectly. We initialised the linear
momentum of the box to zero at the beginning and it remained close to machine
precision throughout the entire simulation.

\section{\label{sec:concl} Conclusions}
In this paper, we have presented a new algorithm  
for NEMD simulations of thermal
gradients. The method comprises an extension to the HEX algorithm, which
rescales and shifts velocities of particles inside reservoirs
to impose a constant heat flux. The problem with the original algorithm is that
it exhibits a drift in the total energy whose origin remained hitherto unclear.
For long simulations, this energy loss becomes restrictive, limiting the
accessible simulation time scales to a few nanoseconds.
In our approach, we reformulated the HEX algorithm as a Trotter factorisation
of the Liouville operator. Using this theoretical framework, it is straightforward to
determine higher-order truncation terms which are a consequence of the employed operator splitting. 
We demonstrated that the leading-order truncation error of the coordinates is
responsible for the observed energy drift.

To test the accuracy of the method, we implemented the eHEX algorithm in
LAMMPS and ran simulations on a Lennard-Jones system and SPC/E water. In
both cases, we observed at least a hundredfold reduction in the energy loss as
compared to the HEX algorithm. With the eHEX algorithm, it is therefore
possible to carry out constant heat flux simulations which are on the order of a hundred nanoseconds and based on fully deterministic equations of
motion.

\begin{acknowledgments}
PW gratefully acknowledges stimulating discussions with Chongli Qin, Clemens
Moritz, Raman Ganti and Aleks Reinhardt, as well as financial support
through a DOC Fellowship of the Austrian Academy of Sciences.
PW would also like to thank members of the LAMMPS mailing list, especially Paul Crozier,
Axel Kohlmeyer and Steve Plimpton, for helpful discussions.
Further financial support from the Federation of Austrian Industry (IV) Carinthia 
and the Austrian Science Fund FWF
within the SFB Vicom (project F41) is acknowledged with gratitude. The results
presented here have been achieved in part using the Vienna Scientific Cluster
(VSC). DF acknowledges support from Engineering and Physical Sciences
 Research Council Programme Grant EP/I001352/1.
Additional data related to this publication are available at the University of 
Cambridge data repository (\href{https://www.repository.cam.ac.uk/handle/1810/250539}{https://www.repository.cam.ac.uk/handle/1810/250539}).
\end{acknowledgments}

\appendix

\section{\label{secapp:hex} Heat exchange algorithm}
\subsection{Exact solution}
We would like to show that the rescaling step
\begin{equation}
\label{eqapp:hex}
{\mathbold v_i}(t) = \xi {\mathbold v}_i(0) +  \big(1-\xi\big) 
{\mathbold v}_{\Gamma}(0)
\end{equation}
 of the HEX algorithm is the exact solution of 
\begin{equation}
\label{eqapp:resc}
{\dot {\mathbold v}}_i = \frac{\mathcal F}{2 \mathcal K} (\mathbold v_i -
{\mathbold v}_\Gamma),
\end{equation}
where $\mathcal K$ is given by
Eq.~(\ref{eq:Ekinnt}) and $\xi$ by Eq.~(\ref{eq:hexxii}).
We will first show that $\mathcal F/2 \mathcal K$ is  
independent of the particle velocities and only a function of time.
This can be seen easily by considering the time evolution of the
internal kinetic energy, which is given by
 \begin{subequations}
 \label{eqapp:dKdt}
 \begin{alignat}{2}
 \frac{\mathrm d{\mathcal K}}{\mathrm dt}   &= && \quad	\sum_{{i \in \gamma}} 
 										m_i \left({\mathbold v}_i - {\mathbold v}_{\Gamma} \right) \cdot 
 										\left(\dot{\mathbold v}_i - \dot{\mathbold v}_\Gamma \right) \\
 &=&& \quad \mathcal F 
 \end{alignat}
 \end{subequations}
 and therefore we can write $\mathcal K(t) = \mathcal K(0) + \mathcal F t$. 
We note that we can exchange the order of taking the time derivative and the summation, because the particle positions are fixed
during this operation.
At the same time, it is easy to see that the centre of mass velocity is 
constant in time since
\begin{alignat}{2}
\label{eqapp:vgamma}
\frac{\mathrm d{\mathbold v_\Gamma}}{\mathrm dt} &= && \quad \frac{1}{m_\Gamma}
\sum_{{i \in \gamma}} m_i \dot {\mathbold v}_i  \\
&= && \quad \frac{1}{m_\Gamma} \sum_{{i
\in \gamma}} m_i \left[
 \frac{\mathcal F}{2 \mathcal K} \left(\mathbold v_i - \mathbold
 v_\Gamma\right) \right]\\
&=&& \quad 0.
\end{alignat}
In order to solve Eq.~(\ref{eqapp:resc}) analytically, it is advantageous to
carry out a variable transformation first. Let us consider the
transformation ${\mathbold v_i} \mapsto {\bar {\mathbold v}_i} =  {\mathbold v_i} -
{\mathbold v_\Gamma}$.
With the definition
\begin{equation}
 \lambda(t) = \frac{\mathcal F}{2 \left[\mathcal K(0) + \mathcal F
t\right]},
\end{equation}
we can express the time evolution of the new velocities as
\begin{equation}
\label{eqapp:hexcont}
{\dot {\bar{\mathbold v}}}_i = \lambda {\bar{\mathbold v}}_i.
\end{equation}
The solution of this equation is then given by
\begin{align}
\label{eqapp:hexex}
{\bar{\mathbold v}}_i(t)= \text{e}^{\int_0^t
\mathrm d t'\ \lambda(t')} {\bar{\mathbold v}}_i(0) =\xi    {\bar{\mathbold v}}_i(0)
.
\end{align}
If we substitute the old variables back, we recover
Eq.~(\ref{eqapp:hex}) which proves the assertion.

\subsection{Splitting error}
In this section, we sketch the derivation of the leading-order error term of
the coordinate integration arising from the operator splitting in the HEX algorithm. To this
end, we evaluate the expression
\begin{align}
\label{eqapp:strang2}
&\mathcal E r_{i,\alpha} = \\
& \Bigg( \frac {1}{12} \big[ iL_2, \left[ iL_2,
iL_1 \right] \big] -\frac {1}{24} \big[ iL_1, \left[ iL_1, iL_2 \right]
\big]\Bigg) r_{i,\alpha} \nonumber
\end{align}
for the operators 
\begin{subequations}
\label{eqapp:hexvcont2}
\begin{align}
iL_1 &= \sum_{j=1}^N \sum_{\beta \in \{x,y,z\}}
  \frac{\eta_{j,\beta}}{m_j} \frac{\partial}{\partial v_{j,\beta}},  \\
iL_2 &=  \sum_{j=1}^N  \sum_{\beta \in \{x,y,z\}} \left[
  \frac{f_{j,\beta}}{m_j} \frac{\partial}{\partial v_{j,\beta}} + v_{j,\beta}
  \frac{\partial}{\partial r_{j,\beta}} \right].
\end{align}
\end{subequations}
For the first term in Eq.~(\ref{eqapp:strang2}) we find
\begin{subequations}
\label{eqapp:commL2commL2L1a}
\begin{align}
 &\big[ iL_2, \left[ iL_2, iL_1 \right] \big] r_{i,\alpha} \nonumber   \\
 & = - 2 \sum_{j=1}^N \sum_{\beta} 
 \bigg[ \frac{f_{j,\beta}}{m_j}     \frac{\partial}{\partial v_{j,\beta}} +
 v_{j,\beta}   \frac{\partial}{\partial r_{j,\beta}} \bigg]
 \frac{\eta_{i,\alpha}}{m_i} \\
 & = - \frac{2}{m_i} \sum_{j \in \gamma_{k(\mathbold r_i)}} \sum_{\beta} 
 \frac{f_{j,\beta}}{m_j}     \frac{\partial \eta_{i,\alpha}}{\partial
 v_{j,\beta}}, 
\end{align}
\end{subequations}
omitting summation bounds for $\beta$ for readibility. In 
the last step we assumed that particles do not cross
reservoir boundaries, in which case $\eta_{i,\alpha}$ depends only
on the velocities of particles within the reservoir $\Gamma_{k(\mathbold r_i)}$.
For the second term in Eq.~(\ref{eqapp:strang2}) we find
\begin{equation}
\label{eqapp:commL1commL1L}
  \big[ iL_1, \left[ iL_1, iL_2 \right] \big] r_{i,\alpha} =  \frac{1}{m_i}
  \sum_{j \in \gamma_{k(\mathbold r_i)}} \sum_{\beta} \frac{\eta_{j,\beta}}{m_j}    
  \frac{\partial \eta_{i,\alpha}}{\partial v_{j,\beta}} 
\end{equation}
and combining the two expressions we get
\begin{equation}
\label{eqapp:epsr2}
\mathcal E r_{i,\alpha} =-\frac {1}{6m_i} 
\sum_{j \in \gamma_{k(\mathbold r_i)}} \sum_{\beta} \frac{1}{m_j}
\bigg(f_{j,\beta} + \frac {\eta_{j,\beta}}{4} \bigg) \frac{\partial
\eta_{i,\alpha}}{\partial v_{j,\beta}}. 
\end{equation}
It is straightforward to compute the derivative
\begin{alignat}{2}
\label{eqapp:detadv}
\frac{\partial
 \eta_{i,\alpha}}{\partial v_{j,\beta}} &=  \frac{m_i \mathcal
 F_{\Gamma_{k(\mathbold r_i)}}}{2 \mathcal K_{\Gamma_{k(\mathbold r_i)}}} \Bigg[ 
 \delta_{\alpha,\beta} \bigg( \delta_{i,j} - \frac{m_j}{m_{\Gamma_{k(\mathbold
 r_i)}}} \bigg) \\
&-\frac{m_j}{\mathcal K_{\Gamma_{k(\mathbold r_i)}}}  \Big(v_{j,\beta} -
v_{\Gamma_{k(\mathbold r_i)}, \beta}\Big) \Big(v_{i,\alpha} -
v_{\Gamma_{k(\mathbold r_i)},\alpha}\Big) \Bigg],
\nonumber
\nonumber
\end{alignat}
where $\delta_{i,j}$ is the Kronecker delta. The final
result, Eq.~(\ref{eq:ehexrj}), is then recovered by substituting the
derivative in Eq.~(\ref{eqapp:epsr2}) with the expression above.
 
\nocite{*}
\bibliography{document}

\begin{thebibliography}{25}%
\makeatletter
\providecommand \@ifxundefined [1]{%
 \@ifx{#1\undefined}
}%
\providecommand \@ifnum [1]{%
 \ifnum #1\expandafter \@firstoftwo
 \else \expandafter \@secondoftwo
 \fi
}%
\providecommand \@ifx [1]{%
 \ifx #1\expandafter \@firstoftwo
 \else \expandafter \@secondoftwo
 \fi
}%
\providecommand \natexlab [1]{#1}%
\providecommand \enquote  [1]{``#1''}%
\providecommand \bibnamefont  [1]{#1}%
\providecommand \bibfnamefont [1]{#1}%
\providecommand \citenamefont [1]{#1}%
\providecommand \href@noop [0]{\@secondoftwo}%
\providecommand \href [0]{\begingroup \@sanitize@url \@href}%
\providecommand \@href[1]{\@@startlink{#1}\@@href}%
\providecommand \@@href[1]{\endgroup#1\@@endlink}%
\providecommand \@sanitize@url [0]{\catcode `\\12\catcode `\$12\catcode
  `\&12\catcode `\#12\catcode `\^12\catcode `\_12\catcode `\%12\relax}%
\providecommand \@@startlink[1]{}%
\providecommand \@@endlink[0]{}%
\providecommand \url  [0]{\begingroup\@sanitize@url \@url }%
\providecommand \@url [1]{\endgroup\@href {#1}{\urlprefix }}%
\providecommand \urlprefix  [0]{URL }%
\providecommand \Eprint [0]{\href }%
\providecommand \doibase [0]{http://dx.doi.org/}%
\providecommand \selectlanguage [0]{\@gobble}%
\providecommand \bibinfo  [0]{\@secondoftwo}%
\providecommand \bibfield  [0]{\@secondoftwo}%
\providecommand \translation [1]{[#1]}%
\providecommand \BibitemOpen [0]{}%
\providecommand \bibitemStop [0]{}%
\providecommand \bibitemNoStop [0]{.\EOS\space}%
\providecommand \EOS [0]{\spacefactor3000\relax}%
\providecommand \BibitemShut  [1]{\csname bibitem#1\endcsname}%
\let\auto@bib@innerbib\@empty
\bibitem [{\citenamefont {Evans}\ and\ \citenamefont
  {Hoover}(1986)}]{Evans1986}%
  \BibitemOpen
  \bibfield  {author} {\bibinfo {author} {\bibfnamefont {D.~J.}\ \bibnamefont
  {Evans}}\ and\ \bibinfo {author} {\bibfnamefont {W.~G.}\ \bibnamefont
  {Hoover}},\ }\href {\doibase 10.1146/annurev.fl.18.010186.001331} {\bibfield
  {journal} {\bibinfo  {journal} {Annu. Rev. Fluid Mech.}\ }\textbf {\bibinfo
  {volume} {18}},\ \bibinfo {pages} {243} (\bibinfo {year} {1986})}\BibitemShut
  {NoStop}%
\bibitem [{\citenamefont {Evans}(1982)}]{Evans1982}%
  \BibitemOpen
  \bibfield  {author} {\bibinfo {author} {\bibfnamefont {D.~J.}\ \bibnamefont
  {Evans}},\ }\href {\doibase 10.1016/0375-9601(82)90748-4} {\bibfield
  {journal} {\bibinfo  {journal} {Phys. Lett. A}\ }\textbf {\bibinfo {volume}
  {91}},\ \bibinfo {pages} {457} (\bibinfo {year} {1982})}\BibitemShut
  {NoStop}%
\bibitem [{\citenamefont {Ashurst}(1976)}]{Ashurst1974}%
  \BibitemOpen
  \bibfield  {author} {\bibinfo {author} {\bibfnamefont {W.~T.}\ \bibnamefont
  {Ashurst}},\ }in\ \href@noop {} {\emph {\bibinfo {booktitle} {Advances in
  Thermal Conductivity: 13th International Conference on Thermal Conductivity,
  Lake Ozark, Nov. 1973, Papers}}}\ (\bibinfo  {publisher} {University of
  Missouri-Rolla, Rolla, MO},\ \bibinfo {year} {1976})\ pp.\ \bibinfo {pages}
  {89--98}\BibitemShut {NoStop}%
\bibitem [{\citenamefont {Baranyai}(1996)}]{Baranyai1996}%
  \BibitemOpen
  \bibfield  {author} {\bibinfo {author} {\bibfnamefont {A.}~\bibnamefont
  {Baranyai}},\ }\href {\doibase 10.1103/PhysRevE.54.6911} {\bibfield
  {journal} {\bibinfo  {journal} {Phys. Rev. E}\ }\textbf {\bibinfo {volume}
  {54}},\ \bibinfo {pages} {6911} (\bibinfo {year} {1996})}\BibitemShut
  {NoStop}%
\bibitem [{\citenamefont {Ciccotti}\ and\ \citenamefont
  {Tenenbaum}(1980)}]{Ciccotti1980}%
  \BibitemOpen
  \bibfield  {author} {\bibinfo {author} {\bibfnamefont {G.}~\bibnamefont
  {Ciccotti}}\ and\ \bibinfo {author} {\bibfnamefont {A.}~\bibnamefont
  {Tenenbaum}},\ }\href {\doibase 10.1007/BF01008518} {\bibfield  {journal}
  {\bibinfo  {journal} {J. Stat. Phys.}\ }\textbf {\bibinfo {volume} {23}},\
  \bibinfo {pages} {767} (\bibinfo {year} {1980})}\BibitemShut {NoStop}%
\bibitem [{\citenamefont {Ikeshoji}\ and\ \citenamefont
  {Hafskjold}(1994)}]{Ikeshoji1994}%
  \BibitemOpen
  \bibfield  {author} {\bibinfo {author} {\bibfnamefont {T.}~\bibnamefont
  {Ikeshoji}}\ and\ \bibinfo {author} {\bibfnamefont {B.}~\bibnamefont
  {Hafskjold}},\ }\href {\doibase 10.1080/00268979400100171} {\bibfield
  {journal} {\bibinfo  {journal} {Mol. Phys.}\ }\textbf {\bibinfo {volume}
  {81}},\ \bibinfo {pages} {251} (\bibinfo {year} {1994})}\BibitemShut
  {NoStop}%
\bibitem [{\citenamefont {M\"{u}ller-Plathe}(1997)}]{Muller-Plathe1997}%
  \BibitemOpen
  \bibfield  {author} {\bibinfo {author} {\bibfnamefont {F.}~\bibnamefont
  {M\"{u}ller-Plathe}},\ }\href {\doibase 10.1063/1.473271} {\bibfield
  {journal} {\bibinfo  {journal} {J. Chem. Phys.}\ }\textbf {\bibinfo {volume}
  {106}},\ \bibinfo {pages} {6082} (\bibinfo {year} {1997})}\BibitemShut
  {NoStop}%
\bibitem [{\citenamefont {Kuang}\ and\ \citenamefont
  {Gezelter}(2010)}]{Kuang2010}%
  \BibitemOpen
  \bibfield  {author} {\bibinfo {author} {\bibfnamefont {S.}~\bibnamefont
  {Kuang}}\ and\ \bibinfo {author} {\bibfnamefont {J.~D.}\ \bibnamefont
  {Gezelter}},\ }\href {\doibase 10.1063/1.3499947} {\bibfield  {journal}
  {\bibinfo  {journal} {J. Chem. Phys.}\ }\textbf {\bibinfo {volume} {133}},\
  \bibinfo {pages} {164101} (\bibinfo {year} {2010})}\BibitemShut {NoStop}%
\bibitem [{\citenamefont {Kuang}\ and\ \citenamefont
  {Gezelter}(2012)}]{Kuang2012}%
  \BibitemOpen
  \bibfield  {author} {\bibinfo {author} {\bibfnamefont {S.}~\bibnamefont
  {Kuang}}\ and\ \bibinfo {author} {\bibfnamefont {J.~D.}\ \bibnamefont
  {Gezelter}},\ }\href {\doibase 10.1080/00268976.2012.680512} {\bibfield
  {journal} {\bibinfo  {journal} {Mol. Phys.}\ }\textbf {\bibinfo {volume}
  {110}},\ \bibinfo {pages} {691} (\bibinfo {year} {2012})}\BibitemShut
  {NoStop}%
\bibitem [{\citenamefont {Hafskjold}, \citenamefont {Ikeshoji},\ and\
  \citenamefont {Ratkje}(1993)}]{Hafskjold1993}%
  \BibitemOpen
  \bibfield  {author} {\bibinfo {author} {\bibfnamefont {B.}~\bibnamefont
  {Hafskjold}}, \bibinfo {author} {\bibfnamefont {T.}~\bibnamefont {Ikeshoji}},
  \ and\ \bibinfo {author} {\bibfnamefont {S.~K.}\ \bibnamefont {Ratkje}},\
  }\href {\doibase 10.1080/00268979300103101} {\bibfield  {journal} {\bibinfo
  {journal} {Mol. Phys.}\ }\textbf {\bibinfo {volume} {80}},\ \bibinfo {pages}
  {1389} (\bibinfo {year} {1993})}\BibitemShut {NoStop}%
\bibitem [{\citenamefont {Bresme}, \citenamefont {Hafskjold},\ and\
  \citenamefont {Wold}(1996)}]{Bresme1996}%
  \BibitemOpen
  \bibfield  {author} {\bibinfo {author} {\bibfnamefont {F.}~\bibnamefont
  {Bresme}}, \bibinfo {author} {\bibfnamefont {B.}~\bibnamefont {Hafskjold}}, \
  and\ \bibinfo {author} {\bibfnamefont {I.}~\bibnamefont {Wold}},\ }\href
  {\doibase 10.1021/jp9512321} {\bibfield  {journal} {\bibinfo  {journal} {J.
  Phys. Chem.}\ }\textbf {\bibinfo {volume} {100}},\ \bibinfo {pages} {1879}
  (\bibinfo {year} {1996})}\BibitemShut {NoStop}%
\bibitem [{\citenamefont {Bugel}\ and\ \citenamefont
  {Galliero}(2008)}]{Bugel2008}%
  \BibitemOpen
  \bibfield  {author} {\bibinfo {author} {\bibfnamefont {M.}~\bibnamefont
  {Bugel}}\ and\ \bibinfo {author} {\bibfnamefont {G.}~\bibnamefont
  {Galliero}},\ }\href {\doibase 10.1016/j.chemphys.2008.06.013} {\bibfield
  {journal} {\bibinfo  {journal} {Chem. Phys.}\ }\textbf {\bibinfo {volume}
  {352}},\ \bibinfo {pages} {249} (\bibinfo {year} {2008})}\BibitemShut
  {NoStop}%
\bibitem [{\citenamefont {Aubry}\ \emph {et~al.}(2004)\citenamefont {Aubry},
  \citenamefont {Bammann}, \citenamefont {Hoyt}, \citenamefont {Jones},
  \citenamefont {Kimmer}, \citenamefont {Klein}, \citenamefont {Wagner},
  \citenamefont {{Webb III}},\ and\ \citenamefont {Zimmerman}}]{Aubry2004}%
  \BibitemOpen
  \bibfield  {author} {\bibinfo {author} {\bibfnamefont {S.}~\bibnamefont
  {Aubry}}, \bibinfo {author} {\bibfnamefont {D.~J.}\ \bibnamefont {Bammann}},
  \bibinfo {author} {\bibfnamefont {J.~J.}\ \bibnamefont {Hoyt}}, \bibinfo
  {author} {\bibfnamefont {R.~E.}\ \bibnamefont {Jones}}, \bibinfo {author}
  {\bibfnamefont {C.~J.}\ \bibnamefont {Kimmer}}, \bibinfo {author}
  {\bibfnamefont {P.~A.}\ \bibnamefont {Klein}}, \bibinfo {author}
  {\bibfnamefont {G.~J.}\ \bibnamefont {Wagner}}, \bibinfo {author}
  {\bibfnamefont {E.~B.}\ \bibnamefont {{Webb III}}}, \ and\ \bibinfo {author}
  {\bibfnamefont {J.~A.}\ \bibnamefont {Zimmerman}},\ }\href@noop {} {\enquote
  {\bibinfo {title} {{A robust, coupled approach for atomistic-continuum
  simulation}},}\ }\bibinfo {type} {Tech. Rep.}\ (\bibinfo  {institution}
  {Sandia National Laboratories},\ \bibinfo {year} {2004})\BibitemShut
  {NoStop}%
\bibitem [{\citenamefont {Plimpton}(1995)}]{Plimpton1995}%
  \BibitemOpen
  \bibfield  {author} {\bibinfo {author} {\bibfnamefont {S.}~\bibnamefont
  {Plimpton}},\ }\href {\doibase 10.1006/jcph.1995.1039} {\bibfield  {journal}
  {\bibinfo  {journal} {J. Comput. Phys.}\ }\textbf {\bibinfo {volume} {117}},\
  \bibinfo {pages} {1} (\bibinfo {year} {1995})}\BibitemShut {NoStop}%
\bibitem [{\citenamefont {Frenkel}\ and\ \citenamefont
  {Smit}(2002)}]{Frenkel2002}%
  \BibitemOpen
  \bibfield  {author} {\bibinfo {author} {\bibfnamefont {D.}~\bibnamefont
  {Frenkel}}\ and\ \bibinfo {author} {\bibfnamefont {B.}~\bibnamefont {Smit}},\
  }\href@noop {} {\emph {\bibinfo {title} {{Understanding Molecular
  Simulation}}}},\ \bibinfo {edition} {2nd}\ ed.\ (\bibinfo  {publisher}
  {Academic Press},\ \bibinfo {address} {San Diego},\ \bibinfo {year}
  {2002})\BibitemShut {NoStop}%
\bibitem [{\citenamefont {Toxvaerd}\ and\ \citenamefont
  {Dyre}(2011)}]{Toxvaerd2011}%
  \BibitemOpen
  \bibfield  {author} {\bibinfo {author} {\bibfnamefont {S.}~\bibnamefont
  {Toxvaerd}}\ and\ \bibinfo {author} {\bibfnamefont {J.~C.}\ \bibnamefont
  {Dyre}},\ }\href {\doibase 10.1063/1.3558787} {\bibfield  {journal} {\bibinfo
   {journal} {J. Chem. Phys.}\ }\textbf {\bibinfo {volume} {134}},\ \bibinfo
  {pages} {081102} (\bibinfo {year} {2011})}\BibitemShut {NoStop}%
\bibitem [{\citenamefont {Nos\'{e}}(1984)}]{Nose1984}%
  \BibitemOpen
  \bibfield  {author} {\bibinfo {author} {\bibfnamefont {S.}~\bibnamefont
  {Nos\'{e}}},\ }\href {\doibase 10.1063/1.447334} {\bibfield  {journal}
  {\bibinfo  {journal} {J. Chem. Phys.}\ }\textbf {\bibinfo {volume} {81}},\
  \bibinfo {pages} {511} (\bibinfo {year} {1984})}\BibitemShut {NoStop}%
\bibitem [{\citenamefont {Hoover}(1985)}]{Hoover1985}%
  \BibitemOpen
  \bibfield  {author} {\bibinfo {author} {\bibfnamefont {W.~G.}\ \bibnamefont
  {Hoover}},\ }\href {\doibase 10.1103/PhysRevA.31.1695} {\bibfield  {journal}
  {\bibinfo  {journal} {Phys. Rev. A}\ }\textbf {\bibinfo {volume} {31}},\
  \bibinfo {pages} {1695} (\bibinfo {year} {1985})}\BibitemShut {NoStop}%
\bibitem [{\citenamefont {Tuckerman}, \citenamefont {Berne},\ and\
  \citenamefont {Martyna}(1992)}]{Tuckerman1992}%
  \BibitemOpen
  \bibfield  {author} {\bibinfo {author} {\bibfnamefont {M.}~\bibnamefont
  {Tuckerman}}, \bibinfo {author} {\bibfnamefont {B.~J.}\ \bibnamefont
  {Berne}}, \ and\ \bibinfo {author} {\bibfnamefont {G.~J.}\ \bibnamefont
  {Martyna}},\ }\href {\doibase 10.1063/1.463137} {\bibfield  {journal}
  {\bibinfo  {journal} {J. Chem. Phys.}\ }\textbf {\bibinfo {volume} {97}},\
  \bibinfo {pages} {1990} (\bibinfo {year} {1992})}\BibitemShut {NoStop}%
\bibitem [{\citenamefont {Strang}(1968)}]{Strang1968}%
  \BibitemOpen
  \bibfield  {author} {\bibinfo {author} {\bibfnamefont {G.}~\bibnamefont
  {Strang}},\ }\href {\doibase 10.1137/0705041} {\bibfield  {journal} {\bibinfo
   {journal} {SIAM J. Numer. Anal.}\ }\textbf {\bibinfo {volume} {5}},\
  \bibinfo {pages} {506} (\bibinfo {year} {1968})}\BibitemShut {NoStop}%
\bibitem [{\citenamefont {LeVeque}\ and\ \citenamefont
  {Oliger}(1983)}]{LeVeque1983}%
  \BibitemOpen
  \bibfield  {author} {\bibinfo {author} {\bibfnamefont {R.~J.}\ \bibnamefont
  {LeVeque}}\ and\ \bibinfo {author} {\bibfnamefont {J.}~\bibnamefont
  {Oliger}},\ }\href {\doibase 10.1090/S0025-5718-1983-0689466-8} {\bibfield
  {journal} {\bibinfo  {journal} {Math. Comput.}\ }\textbf {\bibinfo {volume}
  {40}},\ \bibinfo {pages} {469} (\bibinfo {year} {1983})}\BibitemShut
  {NoStop}%
\bibitem [{\citenamefont {Berendsen}, \citenamefont {Grigera},\ and\
  \citenamefont {Straatsma}(1987)}]{Berendsen1987}%
  \BibitemOpen
  \bibfield  {author} {\bibinfo {author} {\bibfnamefont {H.~J.~C.}\
  \bibnamefont {Berendsen}}, \bibinfo {author} {\bibfnamefont {J.~R.}\
  \bibnamefont {Grigera}}, \ and\ \bibinfo {author} {\bibfnamefont {T.~P.}\
  \bibnamefont {Straatsma}},\ }\href {\doibase 10.1021/j100308a038} {\bibfield
  {journal} {\bibinfo  {journal} {J. Phys. Chem.}\ }\textbf {\bibinfo {volume}
  {91}},\ \bibinfo {pages} {6269} (\bibinfo {year} {1987})}\BibitemShut
  {NoStop}%
\bibitem [{\citenamefont {Ryckaert}, \citenamefont {Ciccotti},\ and\
  \citenamefont {Berendsen}(1977)}]{Ryckaert1977}%
  \BibitemOpen
  \bibfield  {author} {\bibinfo {author} {\bibfnamefont {J.-P.}\ \bibnamefont
  {Ryckaert}}, \bibinfo {author} {\bibfnamefont {G.}~\bibnamefont {Ciccotti}},
  \ and\ \bibinfo {author} {\bibfnamefont {H.~J.~C.}\ \bibnamefont
  {Berendsen}},\ }\href {\doibase 10.1016/0021-9991(77)90098-5} {\bibfield
  {journal} {\bibinfo  {journal} {J. Comput. Phys.}\ }\textbf {\bibinfo
  {volume} {23}},\ \bibinfo {pages} {327} (\bibinfo {year} {1977})}\BibitemShut
  {NoStop}%
\bibitem [{\citenamefont {Andersen}(1983)}]{Andersen1983}%
  \BibitemOpen
  \bibfield  {author} {\bibinfo {author} {\bibfnamefont {H.~C.}\ \bibnamefont
  {Andersen}},\ }\href {\doibase 10.1016/0021-9991(83)90014-1} {\bibfield
  {journal} {\bibinfo  {journal} {J. Comput. Phys.}\ }\textbf {\bibinfo
  {volume} {52}},\ \bibinfo {pages} {24} (\bibinfo {year} {1983})}\BibitemShut
  {NoStop}%
\bibitem [{\citenamefont {Swope}\ \emph {et~al.}(1982)\citenamefont {Swope},
  \citenamefont {Andersen}, \citenamefont {Berens},\ and\ \citenamefont
  {Wilson}}]{Swope1982a}%
  \BibitemOpen
  \bibfield  {author} {\bibinfo {author} {\bibfnamefont {W.~C.}\ \bibnamefont
  {Swope}}, \bibinfo {author} {\bibfnamefont {H.~C.}\ \bibnamefont {Andersen}},
  \bibinfo {author} {\bibfnamefont {P.~H.}\ \bibnamefont {Berens}}, \ and\
  \bibinfo {author} {\bibfnamefont {K.~R.}\ \bibnamefont {Wilson}},\ }\href
  {\doibase 10.1063/1.442716} {\bibfield  {journal} {\bibinfo  {journal} {J.
  Chem. Phys.}\ }\textbf {\bibinfo {volume} {76}},\ \bibinfo {pages} {637}
  (\bibinfo {year} {1982})}\BibitemShut {NoStop}%
\end{thebibliography}%

\end{document}